\documentclass{article}
\usepackage{geometry}
\usepackage[utf8]{inputenc}
\usepackage[T1]{fontenc}
\usepackage{amsmath, amssymb}
\usepackage{graphicx}
\usepackage{tabularx, booktabs}
\usepackage{multirow}
\usepackage{color}
\usepackage{subfig}
\usepackage{hyperref}
\usepackage[hang,flushmargin]{footmisc} 

\begin{document}
\title{A large collection of real-world pediatric sleep studies}

\author{Harlin Lee,\textsuperscript{1} Boyue Li,\textsuperscript{1} Shelly DeForte,\textsuperscript{2} Mark L. Splaingard,\textsuperscript{2}\\
Yungui Huang,\textsuperscript{2} Yuejie Chi,\textsuperscript{1*} Simon L. Linwood\textsuperscript{3*}}
\date{}
\maketitle

\begin{abstract}
	\makeatletter{\renewcommand*{\@makefnmark}{}
		\footnotetext{1. Department of Electrical and Computer Engineering, Carnegie Mellon University, 5000 Forbes Avenue, Pittsburgh PA 15213, USA.  \makeatother}}

	\makeatletter{\renewcommand*{\@makefnmark}{}
			\footnotetext{2. Nationwide Children's Hospital, 700 Children's Drive, Columbus OH 43205, USA.\makeatother}}

	\makeatletter{\renewcommand*{\@makefnmark}{}
			\footnotetext{3. School of Medicine, University of California, Riverside, 92521 Botanic Gardens Drive, Riverside CA 92507, USA.\makeatother}}
			
\makeatletter{\renewcommand*{\@makefnmark}{}
	\footnotetext{{*}Corresponding authors:
		Yuejie Chi (yuejiechi@cmu.edu), Simon Lin Linwood (simon.linwood@ucr.edu)
		}  \makeatother}

Despite being crucial to health and quality of life, sleep---especially pediatric sleep---is not yet well understood. This is exacerbated by lack of access to sufficient pediatric sleep data with clinical annotation. In order to accelerate research on pediatric sleep and its connection to health, we create the Nationwide Children’s Hospital (NCH) Sleep DataBank and publish it at Physionet and the National Sleep Research Resource (NSRR), which is a large sleep data common with physiological data, clinical data, and tools for analyses. The NCH Sleep DataBank consists of 3,984 polysomnography studies and over 5.6 million clinical observations on 3,673 unique patients between 2017 and 2019 at NCH. The novelties of this dataset include: 1) large-scale sleep dataset suitable for discovering new insights via data mining, 2) explicit focus on pediatric patients, 3) gathered in a real-world clinical setting, and 4) the accompanying rich set of clinical data. The NCH Sleep DataBank is a valuable resource for advancing automatic sleep scoring and real-time sleep disorder prediction, among many other potential scientific discoveries. 
\end{abstract}

\section*{Background \& Summary}
Sleep is an active process associated with physiological changes that involve multiple organ systems, and is vital for the maturation and daily functioning of infants, children and adolescents. Consequently, disruption of the complex interplay between sleep and other physiological processes can lead to significant medical consequences \cite{mark_l_splaingard_sleep_2016}. Sleep disorders, like obstructive sleep apnea (OSA) \cite{lumeng_epidemiology_2008, beebe_neuropsychological_2004}, can lead to derangements in function that contribute to significant morbidity and even mortality. Sleep can also be disrupted by many organ-specific diseases like asthma, sickle cell disease, renal failure, or depression that alter the course of a particular medical condition and result in a poorer quality of life. 

Sleep disturbances in children are classified as behavioral insomnias of children, sleep-related breathing disorders, parasomnias, sleep-related movement disorders, circadian rhythm disorders or hypersomnias \cite{american_academy_of_sleep_medicine_international_2014}.  These sleep disorders may be associated with excessive daytime sleepiness (rare in young children), hyperactivity–impaired attention, poor school performance from impaired concentration and vigilance, and behavior problems including irritability.

Sleep problems suffer from under-reporting by parents and under-diagnosis by primary care physicians, but are conservatively estimated to occur in approximately 25\% of healthy children younger than 5 years and in up to 80\% of children with special health care needs. Estimates of prevalence of sleep disorders in children vary more widely for behavioral sleep problems like insomnia than organic sleep problems like OSA. 

While some childhood sleep disorders need only medical history to be properly diagnosed and managed, some infants and children require an analysis of the child actually sleeping, called an overnight sleep study or polysomnography (PSG), to accurately diagnose their sleep-related condition. During an overnight PSG, the sleeping child’s physiological signals are recorded under the direct supervision of specially trained sleep technicians, who attach monitoring sensors to special computer software and adjust them during the night. The technician also provides observations about the child’s sleep that are invaluable in making an accurate diagnosis. Video monitoring is also incorporated into the PSG, allowing review of movements necessary to diagnose nocturnal seizures, which occur in about 20\% of children with epilepsy.  

The physiological data collected during a PSG provide a picture of clinically useful information about different sleep stages, sleep disruption, respiratory status during different sleep stages, leg movements, and changes in cardiac rate and rhythm during sleep.  For instance, episodes of OSA may consist of decreased airflow in spite of normal respiratory effort in thoracic and abdominal belts, changes in electroencephalogram (EEG) pattern called arousals, cardiac deceleration, and oxygen desaturation. These findings may be mild during non-random eye movement (non-REM) sleep but profound during REM sleep. 

Computational algorithms that learn from large amounts of data have seen remarkable success in healthcare, particularly with the proliferation of electronic health records (EHR) and improved sensors. Regrettably, without a curated and comprehensive dataset of substantial size and accessibility, pediatric sleep has not been able to fully benefit from such opportunities yet. As a first step, this data descriptor introduces the Nationwide Children's Hospital (NCH) Sleep DataBank, which has 3,984 pediatric sleep studies on 3,673 unique patients conducted at NCH between 2017 and 2019, along with the patients' longitudinal clinical data. They were gathered in the real-world clinical setting at NCH as opposed to, for example, a controlled clinical trial. The published PSG contain the patient's physiological signals as well as the technician's assessment of the sleep stages and descriptions of additional irregularities \cite{kushida_practice_2005}. The accompanying 5.6 million records of clinical data are extracted from the EHR, and are separated into encounters, medications, measurements (e.g. body mass index), diagnoses, and procedures. The dataset is deposited in the National Sleep Research Resource (NSRR) \cite{zhang2018national} and Physionet \cite{goldberger2000physiobank, physionet_nch}, and can be requested from \url{https://sleepdata.org/datasets/nchsdb} or \url{https://physionet.org/content/nch-sleep}. Accompanying code in Python to assist users in interacting with the dataset is published at \url{https://github.com/liboyue/sleep\_study}.

We expect the NCH Sleep DataBank to be used to study many problems related to pediatric sleep, including but not limited to:
\begin{itemize}
    \item Automatic sleep stage classification, especially algorithms that combine modalities beyond EEG or ECG \cite{grigg-damberger_visual_2007, berry_aasm_2017, berry_aasm_2018, fiorillo_automated_2019, yan_multi-modality_2019}.
    \item Automatic real-time sleep disorder (e.g. OSA) detection \cite{mendonca_review_2018, xie_real-time_2012}.
    \item Diagnosis prediction.
    \item Patient subtyping. There is increasing evidence that many sleep disorders (e.g. insomnia \cite{benjamins_insomnia_2017}) are heterogeneous and have different subtypes. Identifying them can help us understand the disorder better and develop a more tailored course of treatment for different groups of patients.
    \item 	Treatment (e.g. medications and procedures) efficacy analysis.
\end{itemize}

\section*{Methods}
\subsection*{Sleep study data acquisition}
The NCH Sleep DataBank contains sleep studies acquired under standard care at NCH between Dec. 16, 2017 and Dec. 31, 2019 using Natus Sleepworks versions 8 and 9 \cite{noauthor_sleepworks_2017, noauthor_sleepworks_2017-1}. Physiological data collected during an overnight sleep study contain: 
\begin{itemize}
    \item Electroencephalogram (EEG) to identify sleep stages,
    \item Electromyelogram (EMG) of chin activity to help identify the decreased tone seen during REM sleep,
    \item Leg EMG to measure leg movements,
    \item Electrooculogram (EOG) to identify characteristic eye movements seen during REM sleep,
    \item Electrocardiogram (ECG) to monitor cardiac rate and rhythm,
    \item Nasal and oral sensors to measure airflow,
    \item Thoracic and abdominal belts to measure chest and abdominal movements during breathing, which is helpful in demonstrating increased or decreased respiratory effort,
    \item Pulse oximetry to measure blood oxygen saturation,
    \item End-tidal carbon dioxide (CO$_2$) measurement of exhaled air to indirectly measure blood CO$_2$ to assess for hypoventilation.
\end{itemize}

Sleep studies were annotated in real time by technicians at the time of the study, and then were staged and scored by a second technician after the study was completed. Technicians annotated studies using a combination of free-form text entries and selections within Natus Sleepworks. Technicians tried to identify all events of interest, however each technician may have their own style of text annotation. Due to the variability in sleep stages in children, NCH does not use automatic scoring of sleep stages. All sleep stages were manually scored by a technician and then verified or changed by a physician board certified in sleep medicine.

Sleep studies were manually downloaded and converted to EDF+ format between May 2019 and Feb. 2020 using Natus Sleepworks version 9. Any gaps in the time-series data were padded with zeros as part of the conversion. The specific acquisition equipment, setup, and montage all followed standard care protocol at NCH. While changes may have been made to some studies, the NCH protocol for PSG is in accordance with the rules and technical specifications recommended by the American Academy of Sleep Medicine \cite{berry_aasm_2017, berry_aasm_2018}. Standard channel names are used and documented in the header of the EDF files, allowing inference of the montage. 

\subsection*{Patient cohort}
The NCH Sleep DataBank consists of 3,984 sleep studies performed on 3,673 unique patients. Of them, 3,400 patients have one sleep study in the dataset, 238 have two studies, and 35 patients have more than two studies, with a maximum of 5 sleep studies for one patient. In terms of gender distribution, 2,068 patients were male, and 1,604 were female, with one unknown. Table 1 shows the distribution of the unique patients' races, where the majority of the patients were White, and about a fifth were Black or African American. In regards to ethnicity, 186 patients were Hispanic or Latino, 3,446 patients were Not Hispanic or Latino, and 41 had ethnicity of Other, Unknown, or No Information. 

\begin{table}[ht]
\centering
\begin{tabular}{l r r}
\toprule
Race description & Count & Percentage \\ 
\midrule
White & 2,433 & 66.24\% \\
Black or African American & 738 & 20.09\% \\
Multiple races & 277 & 7.54\% \\
Asian & 93 & 2.53\% \\
Others and unknown & 132 & 3.59\% \\
\midrule
Total & 3,673 & 100\% \\
\bottomrule
\end{tabular}
\caption{The distribution of 3,673 unique patients' races.}
\label{tab:race}
\end{table}

\begin{figure}[ht]
        \centering
        \includegraphics[width=0.5\linewidth]{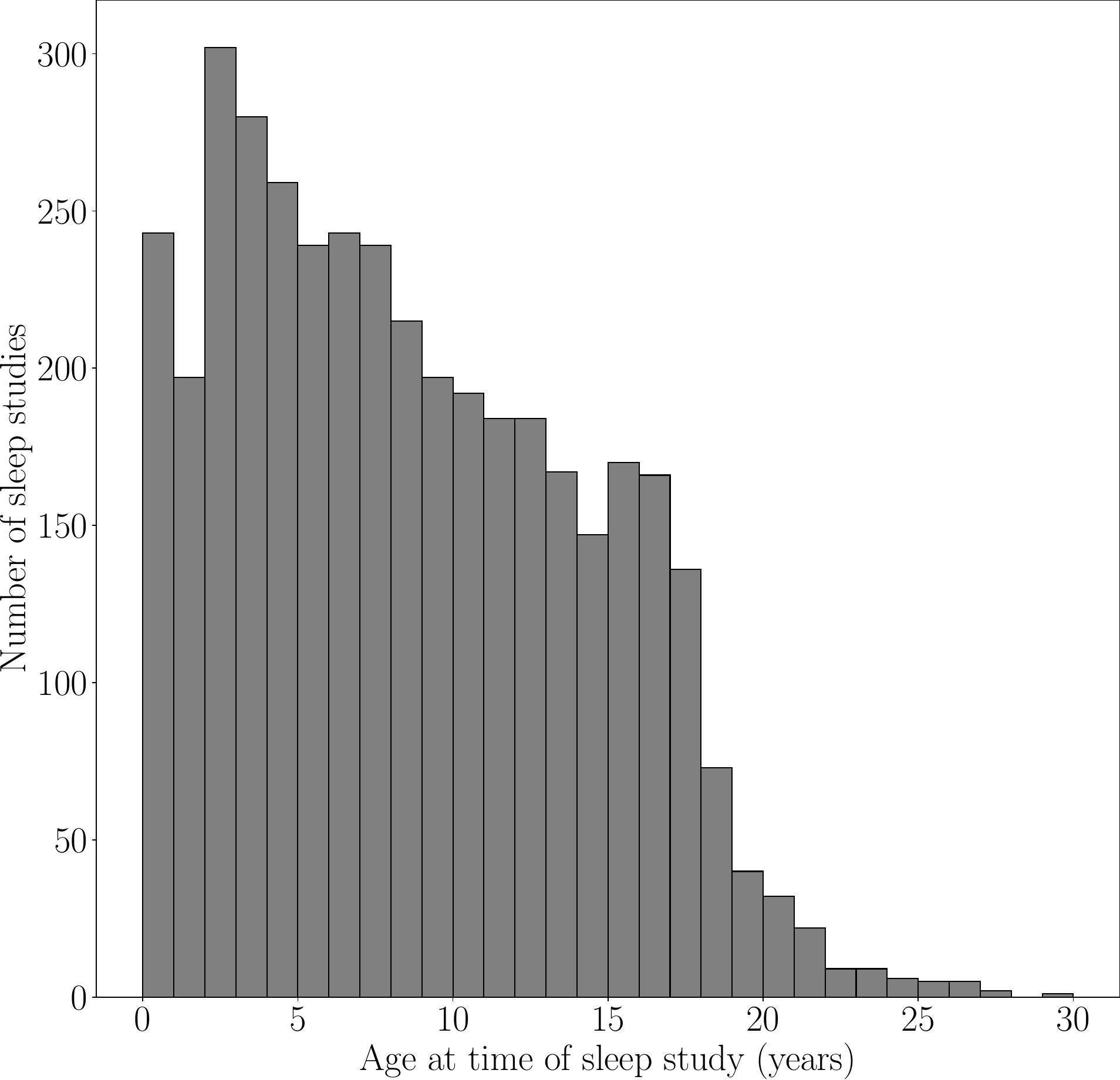}
        \captionof{figure}{Age at the time of sleep study, where $20$ patients that are more than $30$ years old are not shown.}
        \label{fig:age_at_study}
\end{figure}

\begin{figure}[h]
\centering
    \includegraphics[width=\linewidth]{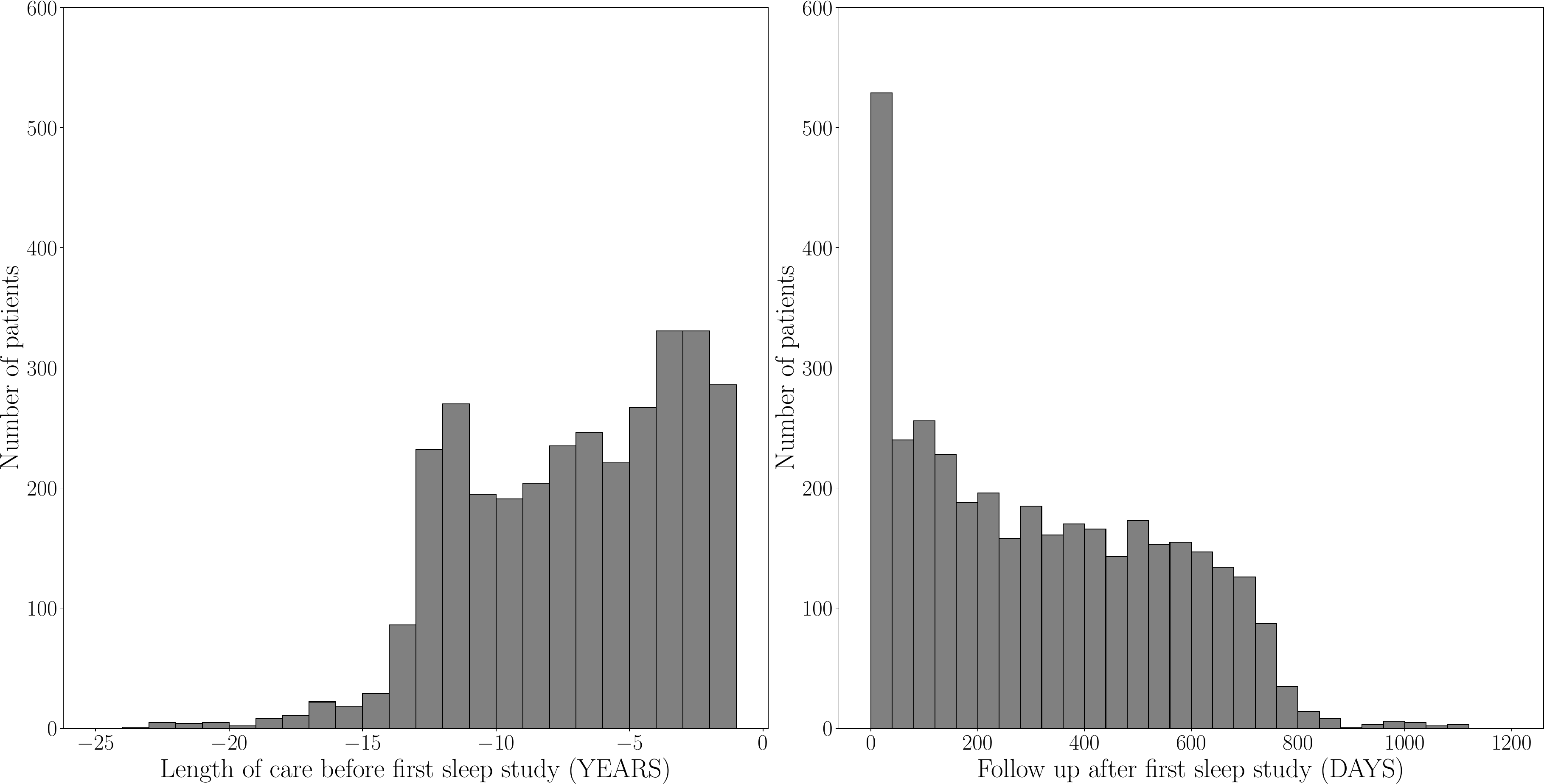}
    \caption{Length of care at NCH before and after first sleep study, where each patient has two entries: one negative for length of care prior to first sleep study (in years), and one positive for follow up after first sleep study (in days). One entry above 1200 days and 5 entries below -25 years are not shown.
  }
    \label{fig:length_of_care_all}
\end{figure}

The majority of patients (2,412) in the dataset were less than 10 years old at the time of the sleep study, as seen in Figure \ref{fig:age_at_study}. Figure \ref{fig:length_of_care_all} summarizes the length of care at NCH before and after the first sleep study. The length of care prior to first sleep study was calculated as the time between the patient’s earliest EHR entry (i.e. diagnosis, encounter, medication, measurement, procedure) and their first sleep study. If the patient’s earliest EHR entry was after the first sleep study, length of care is defined as 0. The length of follow up was calculated as the time between the patient’s first sleep study and their last recorded EHR entry. Patients had a median of 289 days of follow-up after their first sleep study, and 74\% (2,718) had follow-up between 90 days and 2 years.


\subsection*{Patient data linkage}
Sleep study recordings and associated reports at NCH are stored in a database that is independent from the EHR, using Natus Sleepworks as a front end. It was therefore necessary to link patient information in two places. The first link was between the header information in the EDF+ files and the patient data entered in Natus Sleepworks. The second link was between the patient information in Natus and the EHR.

A spreadsheet listing all sleep studies was exported from Natus Sleepworks. This listing included the date and time of each sleep study and patient information such as first and last name, date of birth and medical record number (MRN) for most sleep studies. Sleep studies were then downloaded from Natus in mini-batches, and exported to EDF+. Sleep study specific header information in the EDF+ files were used to match these files to the Natus spreadsheet export. When ambiguity was present, or when MRNs were not present in Natus, we removed the EDF+ file from our dataset. We then used each patient’s last name, date of birth, and MRNs extracted from Natus to retrieve patient records from the EHR. When matches could not be confidently made to the EHR, the sleep studies were removed from the dataset. 

\subsection*{Data de-identification and IRB exemption}
Each unique patient was given a random identifier (STUDY\_PAT\_ID), and each sleep study was given a separate random identifier (SLEEP\_STUDY\_ID). A single patient may have multiple sleep studies in the dataset, and therefore have multiple associated SLEEP\_STUDY\_IDs, but only one STUDY\_PAT\_ID. Sleep studies were then renamed (STUDY\_PAT\_ID)\_(SLEEP\_STUDY\_ID).edf.

All EDF+ headers were de-identified by replacing the first 256 bytes of the EDF+ file with a standard de-identified header. As such, users are advised to ignore all header information in the EDF files (such as patientID, recordID, startdate, duration), but instead rely on the metadata in the accompanying .csv files to interpret the PSG results. Annotation channels were read from EDF+ using Python MNE \cite{gramfort_meg_2013} and written to text. All EDF+ files were converted to EDF by removing the annotation channel using Luna (\url{https://zzz.bwh.harvard.edu/luna}). Annotation text files were then de-identified by replacing any word that was not in a whitelist with ``XXX''. This process affected 10,888 annotations, which is about 0.22\% of the total number of annotations (5,046,370). The whitelist was a combination of 162 common phrases found in the annotations obtained by manual inspection, and a larger whitelist used by the de-identification program Philter \cite{norgeot_protected_2020}. The Philter whitelist contains approximately 195,000 tokens of medical terms and codes and common medical abbreviations, in addition to 20,000 most common English words, and excludes the most common Social Security and Census names. The tab delimited, de-identified annotations were then renamed to match the EDF filenames. 

To protect every patient's privacy, random date shifts were applied to all data: for each patient (i.e. patients with the same STUDY\_PAT\_ID), one random date shift of +/- 180 days was chosen and applied to all data that are linked to the patient.

Finally, we considered the risk of re-identification through rare diagnoses. Say a malicious user of this dataset is interested in re-identifying a specific patient, and the attacker has some information about the patient such as their sex and race, as well as the fact that the patient has been diagnosed for a very rare genetic disease at NCH. If this diagnostic information is visible within the NCH Sleep DataBank, then the malicious user can likely figure out who this patient is, e.g. by searching for a female Asian child born between 2012 and 2016 with this very rare disease. Therefore, we redacted rare diagnosis codes from DIAGNOSIS.CSV through the following procedure as an extra precaution.

The EHR in NCH and the diagnoses table in NCH Sleep DataBank (DIAGNOSIS.CSV) contain several variables. One is DX\_CODE (diagnosis code), which holds the International Classification of Diseases (ICD) code for each diagnosis. On Oct. 1, 2015, hospitals in the United States, including NCH, have switched from using ICD 9 (the 9th revision of ICD) codes to ICD 10 (the 10th revision of ICD) codes in EHR. Another relevant variable is DX\_SOURCE\_TYPE (diagnosis source type), which indicates whether the diagnosis was given at admission, presented as a part of the patient's previous medical history, etc. We were interested in the ones labeled ``Final Dx'', i.e. the final diagnoses the clinicians gave after relevant examinations and tests.

Using these variables, we defined rare diagnosis codes as ICD 10 or ICD 9 codes that were given as Final Dx to less than 10 unique patients in the entire NCH patient population (not limited to the NCH Sleep DataBank patients) during a given time period. Specifically, we queried 1) for every ICD 9 code, the number of unique NCH patients given the diagnosis as Final Dx between Jan. 1, 2000 and Sep. 30, 2015, and 2) for every ICD 10 code, the number of unique NCH patients given the diagnosis as Final Dx between Oct. 1, 2015 and Dec. 31, 2020. If a code had less than 10 unique patients in either ICD 9 or ICD 10 lists, it was deemed a rare diagnosis code for our purpose. We did not consider diagnoses before 2000 since the earliest diagnosis in NCH Sleep DataBank was from 2001.

Then, in every row of DIAGNOSIS.CSV where a rare diagnosis code appeared, we changed the entries in DX\_CODE, DX\_NAME (diagnosis name), DX\_ALT\_CODE (corresponding ICD 10 codes for records before Oct 2015, and ICD 9 codes for those after Oct 2015), CLASS\_OF\_PROBLEM (``Stage 1'', ``Chronic'', ``Acute'', ``Present upon Admission''), CHRONIC\_YN (Indication of chronic disease) to the phrase ``redacted''. This process affected a total of 6,460 rows and 834 unique patients in DIAGNOSIS.CSV.

As this project concerns analysis on de-identified data, the project did not fit the definition of Human Subjects Research as defined by the United States Department of Health and Human Services and Food and Drug Administration. Therefore, this study received NCH Institutional Review Board (IRB) exemption with HIPAA waiver. The protocol that concerns the de-identification and processing of the data, which requires handling identified data, and the collection and publication of data and summary statistics, was approved under “STUDY00000505: Preparation of sleep study data” on September 22, 2019.

\section*{Data Records}

The raw data for NCH Sleep DataBank \cite{physionet_nch} is available at Physionet \url{https://physionet.org/content/nch-sleep}, or at National Sleep Research Resource (NSRR) \url{https://sleepdata.org/datasets/nchsdb}.

The NCH Sleep DataBank consists of two folders: Sleep\_Data and Health\_Data. Sleep\_Data contains annotated PSG recordings, while Health\_Data contains patient demographic and clinical data extracted from the EHR.
Inside Sleep\_Data, PSG sleep studies are provided in the EDF format, and annotations are provided in a separate tab-delimited file. Sleep studies and their matched annotations share the same file name (STUDY\_PAT\_ID)\_(SLEEP\_STUDY\_ID) but different extensions (.edf, .tsv). 
Clinical data in Health\_Data are in .csv files, and they are linked to the files in Sleep\_Data through the same STUDY\_PAT\_ID. Variables follow EHR conventions, and descriptions can be found in the file Sleep\_Study\_Data\_File\_Format.pdf in Health\_Data.

\subsection*{Sleep studies}
The 3,984 sleep study files (.edf) contain PSG recordings taken in clinical setting at NCH. An example plot of the signals can be seen in Figure \ref{fig:ver1}. Almost half (1,972) of the files have 26 channels, a quarter (1,012) have 29, a fifth (820) have 25, and the rest have 28, 24, 40, 27, 9, or 56 channels, in decreasing order of frequency. The most commonly appearing channel names are summarized in Table \ref{tab:common_channels}. The channel PATIENT EVENT was not used and can be excluded from analyses. We note again that all EDF headers were replaced with a standard de-identified version as part of the de-identification process.

The total length of recording in the NCH Sleep DataBank amounts to 40,884 hours, where the minimum length of study is 3 minutes, the maximum is 16.5 hours, and the mean is 10.3 hours. 94.85\% of the files contain between 8 and 12 hours of recordings, and the patients slept for a subset of those times. Users of the dataset should take into account that the majority of the recordings (3,204) are collected with a sampling frequency of 256 Hz, but 581 studies were sampled in 400 Hz, and the rest (199) in 512 Hz.
\begin{table}[]
\centering
    \vspace*{-2cm}
\begin{tabular}{p{13cm}ll}
\toprule
Channel name     \hfill \textit{Description}                      & Count & Percentage \\ \midrule
EEG C3-M2                                      & 3,971            & 99.67\%    \\ \addlinespace[1mm]
EEG O1-M2                                      & 3,971            & 99.67\%    \\\addlinespace[1mm]
EEG O2-M1                                      & 3,971            & 99.67\%    \\\addlinespace[1mm]
EEG CZ-O1                                      & 3,971            & 99.67\%    \\\addlinespace[1mm]
RATE    \hfill \textit{Pulse oximeter signal integrity}                                & 3,970            & 99.65\%    \\\addlinespace[1mm]
ETCO2   \hfill \textit{End tidal CO2}                                       & 3,970            & 99.65\%    \\\addlinespace[1mm]
CAPNO       \hfill \textit{End tidal CO2 waveform}                                   & 3,970            & 99.65\%    \\\addlinespace[1mm]
RESP RATE     \hfill \textit{Respiratory rate}                                 & 3,970            & 99.65\%    \\\addlinespace[1mm]
SPO2 (2,819) or OSAT (1,152)             \hfill \textit{Oxygen saturation}      & 3,970            & 99.65\%    \\\addlinespace[1mm]
EEG F3-M2                                      & 3,969            & 99.62\%    \\\addlinespace[1mm]
RESP THORACIC (2,821) or RESP CHEST (1,148) \hfill \textit{Thoracic inductance}   & 3,969            & 99.62\%    \\\addlinespace[1mm]
RESP ABDOMINAL (2,821) or RESP ABDOMEN (1,148) \hfill \textit{Abdominal inductance} & 3,969            & 99.62\%    \\\addlinespace[1mm]
SNORE      \hfill \textit{Measure of snore or air vibrations}                                    & 3,968            & 99.60\%    \\\addlinespace[1mm]
EEG C4-M1                                      & 3,962            & 99.45\%    \\\addlinespace[1mm]
EEG F4-M1                                      & 3,960            & 99.40\%    \\\addlinespace[1mm]
C-FLOW     \hfill \textit{Continuous positive airflow waveform (PAP only)}                                   & 3,943            & 98.97\%    \\\addlinespace[1mm]
EOG LOC-M2                                     & 3,933            & 98.72\%    \\\addlinespace[1mm]
EOG ROC-M1                                     & 3,931            & 98.67\%    \\\addlinespace[1mm]
EMG CHIN1-CHIN2                                & 3,782            & 94.93\%    \\\addlinespace[1mm]
PRESSURE            \hfill \textit{CPAP pressure (PAP only)}                           & 2,824            & 70.88\%    \\\addlinespace[1mm]
EMG LLEG-RLEG                                  & 2,820            & 70.78\%    \\\addlinespace[1mm]
ECG EKG2-EKG                                   & 2,820            & 70.78\%    \\\addlinespace[1mm]
RESP AIRFLOW        \hfill \textit{Airway pressure with a thermistor}                           & 2,820            & 70.78\%    \\\addlinespace[1mm]
TIDAL VOL         \hfill \textit{Exhaled tidal volume (PAP only)}                             & 2,818            & 70.73\%    \\\addlinespace[1mm]
RESP PTAF       \hfill \textit{Airway pressure with nasal cannula}                             & 2,817            & 70.71\%    \\\addlinespace[1mm]
PATIENT EVENT                         & 2,722            & 68.32\%    \\\addlinespace[1mm]
TCCO2          \hfill \textit{Transcutaneous CO2}                                & 1,417            & 35.57\%    \\\addlinespace[1mm]
SNORE\_DR        \hfill  \textit{Derived snore from PTAF}                              & 1,148            & 28.82\%    \\\addlinespace[1mm]
XFLOW   \hfill \textit{Derived airflow from Resp chest and abdominal}                                     & 1,148            & 28.82\%    \\\addlinespace[1mm]
EMG LLEG+-LLEG-                                & 1,146            & 28.77\%    \\\addlinespace[1mm]
EMG RLEG+-RLEG-                                & 1,146            & 28.77\%    \\\addlinespace[1mm]
ECG LA-RA                                      & 1,146            & 28.77\%    \\\addlinespace[1mm]
FLOW\_DR    \hfill \textit{Derived flow from Resp airflow}                                   & 1,146            & 28.77\%    \\\addlinespace[1mm]
RESP FLOW    \hfill \textit{Airflow channel}                                  & 1,146            & 28.77\%    \\\addlinespace[1mm]
C-PRESSURE    \hfill \textit{Positive pressure delivered via a PAP device}                                 & 1,146            & 28.77\%    \\\addlinespace[1mm]
EEG CHIN1-CHIN2                                & 136              & 3.41\%   \\
\bottomrule
\end{tabular}
    \caption{List of 33 most common channels and their frequencies in 3,984 EDF files. Other 101 channels appear in less than 1\% of the files. Brief descriptions are included for channels that are not measuring EEG, EOG, or EMG. CO2 is carbon dioxide, PAP is positive airway pressure, CPAP is continuous PAP, and PTAF is pressure transducer.}
    \label{tab:common_channels}
\end{table}

\subsection*{Sleep study annotations}
The 3,984 annotation files (.tsv) contain a total of 5,046,370 annotations. The minimum number of annotations contained in a sleep study is 5, while the maximum is 6,047, and the mean value is 1,267. 
Each annotation has the following information, where an example is given in Table \ref{tab:annotation_example}.

\begin{itemize}
\item onset: The start time of the event since the beginning of the study  in seconds.
\item duration: The length of the event in seconds.
\item description: The description of the event, which may be sleep stage label or free-form text entry by the NCH technician, or standard sleep event label by Natus Sleepworks.
\end{itemize}

\begin{table}[]
\centering
    \begin{tabular}{ l r r }
    \toprule
    onset & duration & description  \\
    \midrule
    15985.234375 & 0.0 & Chewing motion \\
    15990.93359375 & 30.0 & Sleep stage W \\
    16002.09375 & 0.0 & Movement  \\
    16002.34375 & 1.21875 & Limb Movement \\
    \bottomrule
    \end{tabular}
    \caption{Example annotations from a .tsv file. ``Chewing motion'' and ``Movement'' are free text entries by the NCH technicion, while ``Limb Movement'' is a standard sleep event labeled by Natus Sleepworks.}
    \label{tab:annotation_example}
\end{table}

35,821 unique descriptions appear in NCH Sleep DataBank. In particular, sleep stages are found in annotations with a duration of 30 seconds, where the descriptions include ``Sleep stage W'', ``Sleep stage N1'', ``Sleep stage N2'', ``Sleep stage N3'', ``Sleep stage R'', or ``Sleep stage ?''. In sleep stage classification, W indicates awake, R stands for REM sleep, and N1, N2, N3 are non-REM stages 1, 2, 3, respectively. The annotation ``Sleep stage ?'' typically occurs after ``Lights On'', and physiological data acquired during that time can usually be ignored, as it indicates that the study has ended. Of the total number of annotations, 79.48\% were related to sleep staging: 6.88\% (347,294) are ``Sleep stage ?'', 13.19\% (665,676) are ``Sleep stage W'', 2.54\% (128,410) are ``Sleep stage N1'', 27.41\% (1,383,765) are ``Sleep stage N2'', 17.35\% (875,486) are ``Sleep stage N3'', and 12.11\% (611,320) are ``Sleep stage R''. This is equivalent to 30,539 hours of data with sleep stage labels. The mean length of such data per study is 7.7 hours, and 96.63\% (3,850) of the studies contain between 6 and 10 hours of sleep data with stage labels.

Besides sleep stage labels, the most common events include: Oxygen Desaturation, Oximeter Event, EEG Arousal, Obstructive Hypopnea, Limb Movement, Gain/Filter Change, Move, Body Position: (Left, Right, Supine, Prone, Upright), Obstructive Apnea, Hypopnea, Central Apnea, and Mixed Apnea.

Free text annotations by the NCH technician typically describe events in the room, movements, and other patient activities, and will often have a duration of 0 seconds. Additionally, hypopneas, apneas, seizures, and other patient events may be mentioned in the free text annotations.
On the other hand, standard sleep event annotations are selected in, or automatically applied by Natus Sleepworks \cite{noauthor_sleepworks_2017, noauthor_sleepworks_2017-1}, and are likely to have varying durations other than 0 or 30 seconds. 

While there may be some variation, the general format for sleep studies is as follows:
Sleep staging begins at the annotation ``Lights Off'' and ends at ``Lights On''. Descriptive annotations will typically precede sleep stage scoring at irregular intervals prior to ``Lights Off''. Sleep stages are annotated in 30 second epochs, beginning at ``Lights Off''; however not all studies include this annotation.

\subsection*{Clinical data}    
The NCH Sleep DataBank includes patient demographics and longitudinal clinical data such as encounters, medication, measurements, diagnoses, and procedures. The number of observations and variables for each file are listed in Table \ref{tab:patient_data}. More details about the variables can be found in Sleep\_Study\_Data\_File\_Format.pdf in the same folder. Note that the age of the patient at the time of sleep study is calculated in SLEEP\_STUDY.csv.  Measurements include body mass index, body mass index percentile, or blood pressure. 

Table \ref{tab:top_diagnoses} lists 20 diagnoses that are given to the highest number of unique patients in the NCH Sleep DataBank according to DIAGNOSIS.csv. Only diagnoses indicated as Final Dx in DX\_SOURCE\_TYPE were considered for this analysis. Any DX\_CODEs recorded in ICD 9 code were converted to the corresponding ICD 10 codes, according to the ICD 10 codes provided under the variable DX\_ALT\_CODE in DIAGNOSIS.csv. 17 unique ICD 9 diagnoses (across 75 rows) that did not have corresponding ICD 10 codes were disregarded from further consideration. We leveraged the hierarchical structure of ICD 10 codes to get a broad overview of the patient population. For example, ICD 10 code ``G47.33 Obstructive sleep apnea (adult) (pediatric)'' fall under the more general ICD 10 code ``G47.3 Sleep apnea'' which in turn is under the even more general ICD 10 code “G47 Sleep disorders.” Therefore, two patients with ``G47.33 Obstructive sleep apnea (adult) (pediatric)'' and ``G47.61 Periodic limb movement disorder'', respectively, counted as two patients diagnosed with “G47 Sleep disorders” in Table \ref{tab:top_diagnoses}. Note that we started by considering all diagnoses in the EHR data, not just the diagnoses resulting from the specific sleep studies included the NCH Sleep DataBank.  

\begin{table}[]
\centering
\begin{tabular}{p{5cm} p{8cm} r}
\toprule
File name      &  Variable names & Rows \\
\midrule
DEMOGRAPHIC.csv      & study pat ID, birth date, pcori gender cd, pcori race cd, pcori hispanic cd, gender descr, race descr, ethnicity descr, language descr, peds gest age num weeks, peds gest age num days
& 3,673\\ \addlinespace[1mm]
SLEEP\_STUDY.csv  &  study pat ID, sleep study ID, sleep study start datetime, sleep study duration datetime, age at sleep study days
  &3,984 \\ \addlinespace[1mm]
SLEEP\_ENC\_ID.csv  & study pat ID, sleep study ID, study enc ID & 3,964 \\ \addlinespace[1mm]
ENCOUNTER.csv       &    study enc ID, study pat ID, encounter date, visit start datetime, visit end datetime, adt arrival datetime, ed departure datetime, encounter type, visit type cd, visit type descr, ICU visit Y/N, prov ID, prov type, dept ID, dept specialty, admit source, hosp admit source, discharge disposition, discharge destination, drg code, drg name, visit reason
  &  495,138         \\ \addlinespace[1mm]
MEDICATION.csv      &  study med ID, study enc ID, study pat ID, med start datetime, med end datetime, med order datetime, med taken datetime, med source type, quantity, days supply, frequency, effective drug dose, eff drug dose source value, drug dose unit, refills, RxNorm code, RxNorm term type, medication descr, generic drug descr, drug order status, drug action, route, route source value, prescribing prov ID, pharm class, pharm subclass, thera class, thera subclass  &   3,035,986        \\ \addlinespace[1mm]
MEASUREMENT.csv            &  study meas ID, study pat ID, study enc ID, meas recorded datetime, meas type, meas value number, meas value text, meas source, study prov ID  & 332,569    \\ \addlinespace[1mm]
DIAGNOSIS.csv             &    study dx ID, study enc ID, study pat ID, dx start datetime, dx end datetime, dx source type, dx enc type, dx code type, dx code, dx name, dx alt code, class of problem, chronic Y/N, prov ID & 1,513,853 \\ \addlinespace[1mm]
PROCEDURE.csv                 &  study proc ID, study pat ID, study enc ID, procedure datetime, study prov ID, proc ID NCH, proc code, proc code type, proc descr & 283,599     \\ \addlinespace[1mm]
PROCEDURE\_SURG\_HX.csv            &   study surghx ID, study pat ID, proc noted date, proc start time, proc end time, proc code, cpt code, proc descr
 & 10,190   \\ \addlinespace[1mm]
\bottomrule
\end{tabular}
\caption{The variable names and number of observations for each patient data file in Health\_Data. More details about the variables can be found in Sleep\_Study\_Data\_File\_Format.pdf in the same folder.
}
\label{tab:patient_data}
\end{table}

\begin{table}[]
\centering
\begin{tabular}{p{8cm} r r}
\toprule
Diagnosis                                                                                           & ICD 10 code & Patients, $N$ \\ \midrule
Sleep disorders                                                                                              & G47 \qquad & 3,379                \\\addlinespace[1mm]
\qquad Sleep   apnea                                                                                                & G47.3 & 2,558                \\
\qquad Sleep   disorder, unspecified                                                                                & G47.9 & 1,163                \\
\qquad Other   sleep disorders                                                                                      & G47.8 & 914                  \\
\qquad Circadian   rhythm sleep disorders                                                                           & G47.2 & 566                  \\
\qquad Insomnia                                                                                                     & G47.0 & 388                  \\
\qquad Hypersomnia                                                                                                  & G47.1 & 257                  \\
\qquad Sleep   related movement disorders                                                                           & G47.6 & 180                  \\
\qquad Parasomnia                                                                                                   & G47.5 & 165                  \\
\qquad Narcolepsy   and cataplexy                                                                                   & G47.4 & 47                   \\\addlinespace[1mm]
Abnormalities of breathing                                                                                   & R06 \qquad & 2,776                \\\addlinespace[1mm]
Encounter for immunization                                                                                   & Z23 \qquad & 1,720                \\\addlinespace[1mm]
Chronic diseases of tonsils and adenoids                                                                     & J35 \qquad & 1,686                \\\addlinespace[1mm]
Encounter for general examination without   complaint, suspected or reported diagnosis                       & Z00 \qquad & 1,587                \\\addlinespace[1mm]
Acute upper respiratory infections of multiple   and unspecified sites                                       & J06 \qquad & 1,537                \\\addlinespace[1mm]
Body mass index (BMI)                                                                                        & Z68 \qquad & 1,417                \\\addlinespace[1mm]
Suppurative and unspecified otitis media                                                                     & H66 \qquad & 1,378                \\\addlinespace[1mm]
Symptoms and signs concerning food and fluid   intake                                                        & R63 \qquad & 1,369                \\\addlinespace[1mm]
Acute pharyngitis                                                                                            & J02 \qquad & 1,260                \\\addlinespace[1mm]
Other symptoms and signs involving the   circulatory and respiratory system                                  & R09 \qquad & 1,256                \\\addlinespace[1mm]
Other functional intestinal disorders                                                                        & K59 \qquad & 1,185                \\\addlinespace[1mm]
Cough                                                                                                        & R05 \qquad & 1,176                \\\addlinespace[1mm]
Lack of expected normal physiological development   in childhood and adults                                  & R62 \qquad & 1,097                \\\addlinespace[1mm]
Encounter for follow-up examination after completed   treatment for conditions other than malignant neoplasm & Z09 \qquad & 1,068                \\\addlinespace[1mm]
Nausea and vomiting                                                                                          & R11 \qquad & 1,051                \\\addlinespace[1mm]
Fever of other and unknown origin                                                                            & R50 \qquad & 1,043                \\\addlinespace[1mm]
Specific developmental disorders of speech and   language                                                    & F80 \qquad & 1,002                \\\addlinespace[1mm]
Asthma                                                                                                       & J45 \qquad & 991                  \\\addlinespace[1mm]
Gastro-esophageal reflux disease                                                                             & K21 \qquad & 982                  \\\addlinespace[1mm]
 \bottomrule 
\end{tabular}
\caption{20 diagnoses that are given to the highest number of unique patients in the NCH Sleep DataBank according to DIAGNOSIS.csv. Note that the diagnoses were abstracted to a higher level before being counted. For example, patients with diagnosis “G47.33 Obstructive sleep apnea (adult) (pediatric)” were counted under G47 and G47.3.}
 \label{tab:top_diagnoses}
\end{table}

\section*{Technical Validation}

\subsection*{Validation of de-identification procedure}
After EDF files were de-identified, we performed several validation steps to confirm that the data matched the original EDF+ export. We loaded all channels from both the de-identified EDF file and the original EDF+ export and confirmed that all signal channels matched. 
Finally, all files included in the data set have been read by Python MNE through this validation procedure and any files with read errors were not included in the data set.

\subsection*{Validation of data maps}
We identified and tested three separate points in our data pipeline: 1) mapping of sleep study from Natus Sleepworks to the de-identified EDF file, 2) mapping of clinical data from EHR to the de-identified CSV files, and 3) the linkage between the sleep study and the clinical data. 

\begin{figure}[]
    \centering
    \vspace*{-3.5cm}
    \centerline{\hspace{-0.5cm}
 \includegraphics[width=1.24\linewidth]{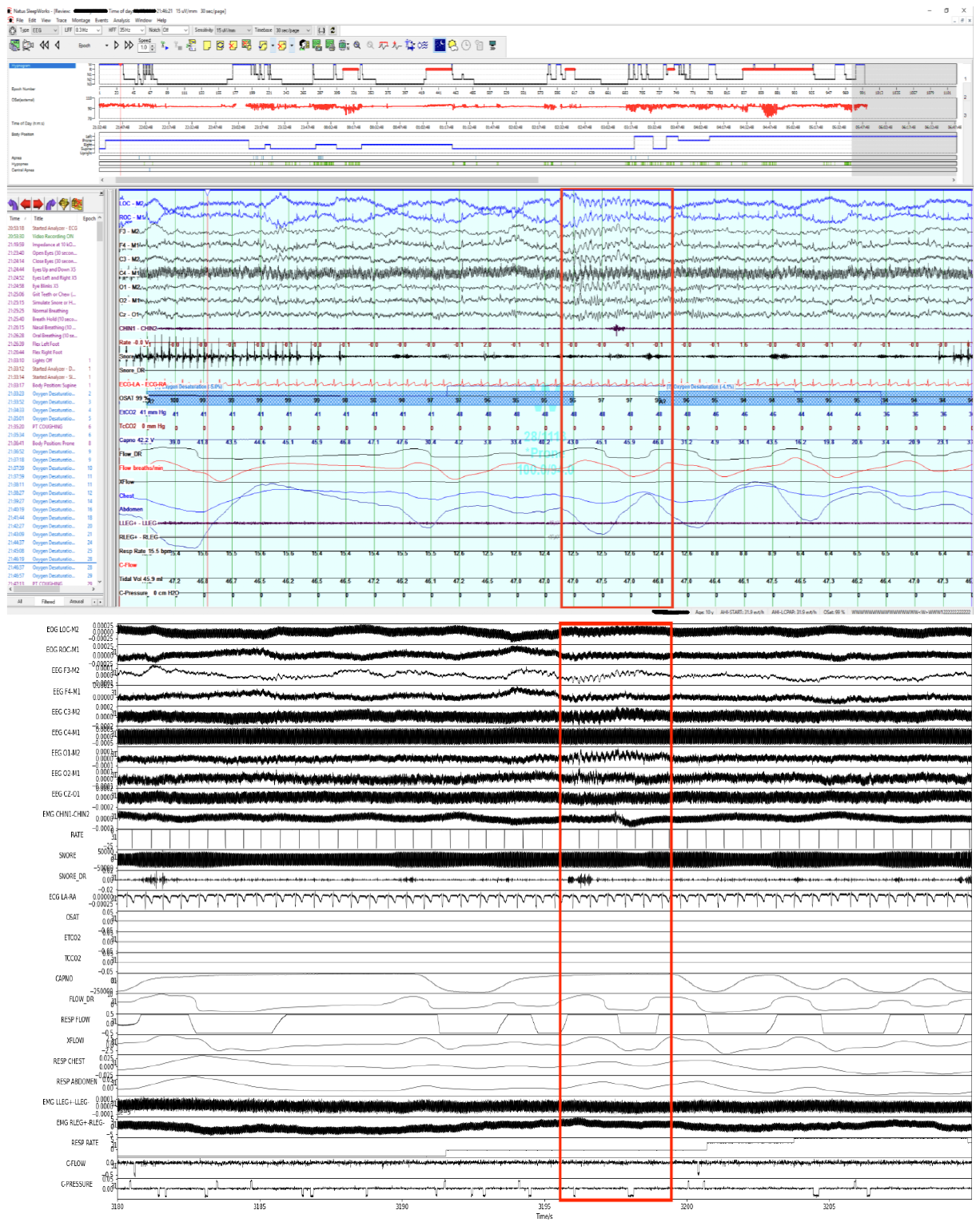}}
    \caption{Visual verification that a randomly chosen 30-second segment of sleep data on Natus Sleepworks (top) matches the sleep data in the corresponding EDF file (bottom), especially at the region of interest marked by red box. Natus Sleepworks may denoise or auto-scale some signals for the viewer.}
    \label{fig:ver1}
\end{figure}

The first was the mapping between the de-identified EDF file and the original sleep data file accessible via Natus Sleepworks. We first chose four random sleep studies (about 0.1\% of the dataset), and a random 30-second segment from each study. Then we confirmed that the sleep data viewed on Natus Sleepworks (Figure \ref{fig:ver1} top) matched data visualized from the corresponding EDF file in the published dataset (Figure \ref{fig:ver1} bottom).

The second mapping was between the de-identified clinical data and the EHR. We extracted from the dataset all clinical data associated with the four random patients chosen in the first verification step, and confirmed that they are identical to the medical records viewed from the physician interface of the EPIC electronic medical record.

The last mapping we verified was SLEEP\_STUDY\_ID, the random identifier linking the sleep studies to the patient data. We verified this by matching the sleep study, which is represented by SLEEP\_STUDY\_ID, with its corresponding encounter in the patient data, which is represented by STUDY\_ENC\_ID. If an encounter had procedure codes and departmental codes associated with sleep study, had the same randomly assigned STUDY\_PAT\_ID as the sleep study, and the same starting date and time (within a window of +/- one hour) as the sleep study start time obtained from Natus Sleepworks, we considered it a match. We were able to match 3,964 sleep studies to encounter codes in the patient data using this method, therefore providing validation of a mapping between the sleep studies and patient data and consistency of date shifting. This information is provided in the file SLEEP\_ENC\_ID.csv.

\subsection*{Sleep stage classification for PSG data validation} 

We developed a baseline sleep stage classifier and included it in the codebase to demonstrate the technical quality as well as a potential utility of the dataset, especially the PSG data. This simple algorithm predicts the sleep stages (W, N1, N2, N3, R) based on 30 seconds of 7 EEG channels (F4-M1, O2-M1, C4-M1, O1-M2, F3-M2, C3-M2, CZ-O1) after they are down sampled to 128Hz. 

Wavelet transform is a powerful method that can flexibly represent the time-frequency content of a signal. As such, it is particularly useful in analyzing non-stationary signals, and have previously been used for EEG-based sleep stage classification \cite{ebrahimi_automatic_2008, fraiwan_automated_2012, hassan_decision_2016, sen_comparative_2014}. After applying multi-resolution Daubechies wavelet transform \cite{daubechies_orthonormal_1988} to each EEG channel, we computed summary statistics such as min, max, mean, and standard deviation of the coefficients, resulting in 84 features. A random forest classifier with 100 decision trees was then trained on these features using 67\% of the dataset, and tested on the rest. 

Table \ref{tab:sleep_scoring} reports the 3-fold stratified cross validation results on 3,928 sleep studies that had the 7 EEG channels, in addition to the results on some subgroups (0 to 1 year old, 1 to 2 years old, and 18+ patients). Fitting the classifier with default parameters from Scikit-learn \cite{scikit-learn} took ~1 hour on Intel Xeon Gold 3.60GHz CPU in parallel; subgroups took less than 2 minutes each. This quick and straightforward algorithm, without any denoising or parameter tuning, achieves a classification accuracy of over 80\% on the 222 adult sleep studies, suggesting high quality of the PSG recordings. Moreover, the difference in classification results between age groups supports the importance of having a dataset dedicated to pediatric sleep.

\begin{table}[]
\centering
\subfloat[All age groups. 3,928 sleep studies and 3,644,305 samples. Overall accuracy is 64.4\%.]{
\begin{tabular}{rrrrrrr}
\toprule
\multicolumn{2}{l}{\multirow{2}{*}{}}                            & \multicolumn{5}{r}{Automated   score sleep stage} \\
\multicolumn{2}{l}{}                                             & W        & N1      & N2       & N3      & R       \\\midrule
\multirow{5}{*}{\rotatebox{90}{\parbox{23mm}{\raggedright Manual   score sleep stage, $N$}}} & W (661,645)    & \textbf{63.1}     & 0.      & 34.0     & 1.5     & 1.4     \\\addlinespace[1mm]
                                                & N1 (127,602)   & 23.9     & \textbf{0.9}     & 68.1     & 2.1     & 5.0     \\\addlinespace[1mm]
                                                & N2 (1,375,678) & 4.4      & 0.      & \textbf{88.6}     & 5.8     & 1.1     \\\addlinespace[1mm]
                                                & N3 (871,200)   & 1.7      & 0.      & 27.2     & \textbf{70.7}    & 0.      \\\addlinespace[1mm]
                                                & R (608,180)    & 6.7      & 0.      & 76.6     & 1.5     & \textbf{15.1}  \\ \bottomrule
\end{tabular}
}\quad%
    \subfloat[18 years and older. 222 sleep studies and 196,135 samples.  Overall accuracy is 81.1\%.]{ 
\begin{tabular}{rrrrrrr}
\toprule
\multicolumn{2}{l}{\multirow{2}{*}{}}                         & \multicolumn{5}{r}{Automated   score sleep stage} \\
\multicolumn{2}{l}{}                                          & W        & N1      & N2       & N3      & R       \\ \midrule
\multirow{5}{*}{\rotatebox{90}{\parbox{23mm}{\raggedright Manual   score sleep stage, $N$}}} & W (52,979)  & \textbf{89.5}     & 0.1     & 8.2      & 0.5     & 1.7     \\ \addlinespace[1mm]
                                                & N1 (8,263)  & 37.5     & \textbf{2.5}     & 47.4     & 0.6     & 12.1    \\ \addlinespace[1mm]
                                                & N2 (80,275) & 5.6      & 0.1     & \textbf{89.1}     & 2.9     & 2.3     \\ \addlinespace[1mm]
                                                & N3 (30,612) & 2.6      & 0.      & 18.3     & \textbf{79.1}    & 0.      \\ \addlinespace[1mm]
                                                & R (24,006)  & 9.2      & 0.      & 24.7     & 0.6     & \textbf{65.5}   \\ 
                                          \bottomrule       
\end{tabular}}\quad%
\subfloat[0-1 year olds. 242 sleep studies and 230,824 samples. Overall accuracy is 76.6\%.]{  
\begin{tabular}{rrrrrrr}
\toprule
\multicolumn{2}{l}{\multirow{2}{*}{}}                         & \multicolumn{5}{r}{Automated score sleep stage} \\
\multicolumn{2}{l}{}                                          & W        & N1     & N2      & N3      & R       \\\midrule
\multirow{5}{*}{\rotatebox{90}{\parbox{23mm}{\raggedright Manual   score sleep stage, $N$}}} & W (63,041)  & \textbf{83.3}     & 0.     & 2.4     & 2.8     & 11.4    \\ \addlinespace[1mm]
                                                & N1 (4,579)  & 28.7     & \textbf{1.1}   & 24.7    & 6.2     & 39.2    \\ \addlinespace[1mm]
                                                & N2 (38,525) & 9.4      & 0.     & \textbf{62.9}    & 10.2    & 17.4    \\ \addlinespace[1mm]
                                                & N3 (64,512) & 4.5      & 0.     & 3.7     & \textbf{83.3}    & 8.5     \\ \addlinespace[1mm]
                                                & R (60,167)  & 11.1     & 0.     & 5.0     & 7.1     & \textbf{76.8}   \\
                                                 \bottomrule
\end{tabular}
}
\caption{Sleep stage classification results of our baseline algorithm applied to different age groups. One sample is a 30-second epoch of sleep. Cell (row $i$, column $j$) of the normalized confusion matrix indicates the percentage (\%) of samples in stage $i$ (manually scored by NCH technician) that were predicted to be in stage $j$ (by our automated algorithm). Each row adds to 100\%. Bolded diagonal entries are the percentages of samples in each stage that were correctly classified. Overall accuracy is the total number of correctly classified samples divided by the total number of samples in \%. All numbers reported are averaged over 3-fold stratified cross validation trials and rounded to one decimal point. Standard deviation was <1\% for all entries except one and not shown here.
}
\label{tab:sleep_scoring}
\end{table}

\subsection*{Prader-Willi syndrome (PWS) patient analysis for EHR data validation}
The availability of EHR allows the study of clinically meaningful patient subpopulations in the NCH Sleep DataBank. As a use case, we examine the sleep patterns of PWS patients within this dataset. To provide context, PWS is a rare genetic disorder that is estimated to affect 1 out of 10,000 to 30,000 people, and many researchers and clinicians are interested in sleep abnormalities and sleep-disordered breathing of PWS patients  \cite{vela_sleep_1984, hertz_sleep_1993, nixon_sleep_2002, meyer_outcomes_2012, pavone_sleep_2015}. We construct two PSG cohorts, where Cohort 1 includes the PSGs of PWS patients, and Cohort 2 consists of PSGs of obese but non-PWS patients. To control for the effect of OSA, both cohorts only consider PSGs during which patients were diagnosed OSA.

To construct the PSG cohorts, we first searched for all STUDY\_ENC\_IDs in DIAGNOSIS.csv during which a patient was given a final diagnosis of OSA. Then, we only kept the encounter IDs that were also present in SLEEP\_ENC\_ID.csv, as we have matched them with SLEEP\_STUDY\_IDs in an earlier validation step. This process identified 860 PSGs (763 unique patients) with OSA diagnoses. Among these, 16 PSGs (12 unique patients) were designated Cohort 1, since they were associated with STUDY\_PAT\_IDs that had a final diagnosis of PWS in the EHR. For reference, the NCH Sleep DataBank has a total of 34 unique patients who had final diagnosis of PWS in the EHR. On the other hand, 370 PSGs (311 unique patients) were associated with STUDY\_PAT\_IDs with obesity diagnoses but not PWS, and selected Cohort 2.

For every PSG in Cohort 1 and Cohort 2, we tallied the number of each sleep stage (W, N1, N2, N3, R) annotation, and extracted the following sleep characteristics: total length of sleep (sleep time) by counting 30 seconds of sleep for each sleep stage annotation, and distribution of sleep stages, e.g., W constitutes 20\% of the sleep time. Table \ref{tab:pws} describes summary statistics of the two cohorts’ sleep characteristics. In summary, the ease-of-navigation of the EHR data makes it possible to conduct  disease-specific data mining using NCH Sleep DataBank, e.g. extraction of  sleep characteristics such as apnea-hypopnea index (AHI), and refined statistical analysis that accounts for potential confounding variables such as BMI and age.

\begin{table}[]
\centering
\begin{tabular}{lrr}
\toprule
                                   & Cohort 1       & Cohort 2       \\ \midrule
PSG,   $N$                           & 16             & 370             \\ \addlinespace[1mm]
Unique   patients, $N$               & 12             & 311             \\ \addlinespace[1mm]
Age,   mean $\pm$   s.d. (years)        & 10.5   $\pm$   5.6 & 13.2   $\pm$   4.7  \\ \addlinespace[1mm]
Sleep   time, mean $\pm$   s.d. (hours) & 8.0   $\pm$   0.7  & 7.5   $\pm$   0.9   \\ \addlinespace[1mm]
\qquad W,   mean $\pm$   s.d. (\%)             & 14.4   $\pm$   7.1 & 20.5   $\pm$   16.1 \\ \addlinespace[1mm]
\qquad N1,   mean $\pm$   s.d. (\%)            & 4.1   $\pm$   2.7  & 3.5   $\pm$   3.4   \\ \addlinespace[1mm]
\qquad N2,   mean $\pm$   s.d. (\%)            & 45.2   $\pm$   7.3 & 39.9   $\pm$   11.5 \\ \addlinespace[1mm]
\qquad N3,   mean $\pm$   s.d. (\%)            & 20.5   $\pm$   6.7 & 21.1   $\pm$   8.5  \\ \addlinespace[1mm]
\qquad R,   mean $\pm$   s.d. (\%)             & 15.8   $\pm$   6.0 & 15.0   $\pm$   7.3  \\ \addlinespace[1mm]
\qquad N1   N2, mean $\pm$   s.d. (\%)         & 49.3   $\pm$   6.7 & 43.4   $\pm$   11.8 \\ \addlinespace[1mm]
\qquad N1   N2 N3, mean $\pm$ s.d. (\%)        & 69.8   $\pm$   6.3 & 64.5   $\pm$   13.4
\\ \addlinespace[1mm] \bottomrule
\end{tabular}
\caption{Summary statistics of sleep time and distribution of sleep stages for two PSG cohorts. Cohort 1: PSGs with OSA diagnoses on PWS patients, Cohort 2: PSGs with OSA diagnoses on obese but not PWS patients; sleep time: total amount of time spent in sleep stages W, N1, N2, N3, and R; s.d.: standard deviation. Percentage of each sleep stage is calculated by dividing time spent in each sleep stage by sleep time. All numbers are rounded to one decimal point.}
\label{tab:pws}
\end{table}

\section*{Usage Notes}

The NCH Sleep DataBank can potentially be used to study many problems related to pediatric sleep, including but not limited to:
\begin{itemize}
    \item 	Automatic sleep scoring (sleep stage classification): Sleep scoring divides sleep into two stages, rapid eye movement (REM), and non-REM, then further divides the latter into shallow sleep (stages N1 and N2) and deep sleep (stage N3) \cite{grigg-damberger_visual_2007, berry_aasm_2017, berry_aasm_2018}, in addition to wake (Stage W). In typical pediatric clinical settings, this is a time-consuming and tedious process done by a technician. Many computational algorithms have shown promise for automatic sleep scoring in adults \cite{fiorillo_automated_2019}, which encourage exploration on automatic sleep scoring for infants and children. Algorithms that combine PSG modalities beyond EEG or ECG \cite{yan_multi-modality_2019} especially warrant more investigation.
    \item Automatic sleep disorder (e.g. obstructive apnea) detection: Large sets of PSG signals published with expert annotations can be leveraged to develop computational algorithms in sleep disorder detection, unleashing the potential of eventual real-time systems that read these signals and detect sleep disorders at their onsets \cite{mendonca_review_2018, xie_real-time_2012}. OSA detection is particularly important, as OSA is associated with various cardiovascular, respiratory, and neurocognitive deficits and morbidity among infants and children \cite{lumeng_epidemiology_2008, beebe_neuropsychological_2004}. 
    \item Diagnosis prediction: Statistical models that predict or measure the risk of diagnoses using other variables (e.g. other diagnoses, demographic, features from PSG, encounters, measurement values) can be constructed and validated to create hypotheses for further experiment.
    \item Identifying patient subgroups: Given the demographics and medical history, patients can be divided into clinically meaningful subgroups before further analysis, as demonstrated in this paper for PWS. Additionally, data-driven approaches may be developed to reveal clusters within the patient population, which could affect their symptoms or best courses of treatment, e.g. as suggested for insomnia \cite{benjamins_insomnia_2017}.
    \item Treatment efficacy analysis: Retrospective studies using the accompanying longitudinal clinical data (e.g. medications and procedures) can be used to analyze efficacy of different treatments options.
\end{itemize}

\section*{Competing interests}
The authors declare no competing interests.

\section*{Acknowledgements}

Research reported in this publication was supported by the National Institute Of Biomedical Imaging And Bioengineering of the National Institutes of Health under Award Number R01EB025018. The content is solely the responsibility of the authors and does not necessarily represent the official views of the National Institutes of Health. The authors thank Tim Held for data identification, Melody Kitzmiller for data query, Dan Digby for data pipelines, Rajesh Ganta for data validation, Iris Karhoff for the interpretation of PSG channel names, Rahul Ragesh, Ramachandra Mannava, and Jacob Hoffman for help with sleep stage classifier development, Daniel Mobley and Michael Rueschman for publishing the data to NSRR, and Lucas McCullum and Tom Polland for publishing the data to Physionet.

\section*{Author contributions}

Y.C. and S.L.L. designed and supervised the study. S.D., Y.H., B.L., and H.L. prepared the dataset. M.L.S. provided clinical interpretations. H.L., B.L., and S.D. conducted data analysis and technical validation. H.L., Y.C., S.D., M.S., Y.H., B.L., and S.L.L. drafted the manuscript.

\section*{Code availability}
The code that was used to analyze patient data, read EDF files, run baseline sleep stage classifier, and generate figures and tables in this paper is published at \url{https://github.com/liboyue/sleep_study}.

\bibliographystyle{naturemag}

\end{document}